\documentclass[%
reprint,
twocolumn,
superscriptaddress,
 amsmath,amssymb,
 aps,
 rc ]{revtex4-2}
\usepackage{amsmath,amsthm,amsfonts,amssymb,amscd,mathrsfs,mathtools,physics,esint,cancel,minted,bm}
\usepackage{comment}
\usepackage{graphicx}
\usepackage{dcolumn}
\usepackage{bm}
\usepackage{xcolor}
\usepackage{siunitx}
\usepackage[bookmarks=false,draft]{hyperref}
\usepackage{float}

\begin{document}

\preprint{APS/123-QED}

\title{Laser-induced spectral diffusion of T centers in silicon nanophotonic devices
}

\author{Xueyue Zhang}
\email{Current address: Department of Applied Physics and Applied Mathematics, Columbia University, New York, New York 10027, USA}
\affiliation{Department of Electrical Engineering and Computer Sciences, University of California, Berkeley, Berkeley, California 94720, USA}
\affiliation{Department of Physics, University of California, Berkeley, Berkeley, California 94720, USA}

\author{Niccolo Fiaschi}
\affiliation{Accelerator Technology and Applied Physics Division, Lawrence Berkeley National Laboratory, Berkeley, CA, USA.}
\affiliation{Department of Electrical Engineering and Computer Sciences, University of California, Berkeley, Berkeley, California 94720, USA}
\affiliation{Department of Physics, University of California, Berkeley, Berkeley, California 94720, USA}

\author{Lukasz Komza}
\affiliation{Department of Physics, University of California, Berkeley, Berkeley, California 94720, USA}
\affiliation{Materials Sciences Division, Lawrence Berkeley National Laboratory, Berkeley, California 94720, USA}

\author{Hanbin Song}
\affiliation{Materials Sciences Division, Lawrence Berkeley National Laboratory, Berkeley, California 94720, USA}
\affiliation{
Department of Materials Science and Engineering, University of California, Berkeley, Berkeley, California 94720, USA
}

\author{Thomas Schenkel}
\affiliation{Accelerator Technology and Applied Physics Division, Lawrence Berkeley National Laboratory, Berkeley, CA, USA.}

\author{Alp Sipahigil}
\email{Corresponding author: alp@berkeley.edu}
\affiliation{Department of Electrical Engineering and Computer Sciences, University of California, Berkeley, Berkeley, California 94720, USA}
\affiliation{Materials Sciences Division, Lawrence Berkeley National Laboratory, Berkeley, California 94720, USA}
\affiliation{Department of Physics, University of California, Berkeley, Berkeley, California 94720, USA}

\date{\today}

\begin{abstract}
Color centers in silicon are emerging as spin-photon interfaces operating at telecommunication wavelengths.   
The nanophotonic device integration of silicon color centers via ion implantation leads to significant optical linewidth broadening, which makes indistinguishable photon generation challenging.
Here, we study the optical spectral diffusion of T centers in a silicon photonic crystal cavity. 
We investigate the linewidth broadening  timescales and origins by measuring the temporal correlations of the resonance frequency under different conditions.
Spectral hole burning measurements reveal no spectral broadening at short timescales from 102~ns to 725~ns. 
We probe broadening at longer timescales using a check pulse to herald the T center frequency and a probe pulse to measure frequency after a wait time. The optical resonance frequency is stable up to 3~ms in the dark. Laser pulses below the silicon band gap applied during the wait time leads to linewidth broadening. Our observations establish laser-induced processes as the dominant spectral diffusion mechanism for T centers in devices, and inform materials and feedback strategies for indistinguishable photon generation. 
\end{abstract}

\maketitle

Color centers in silicon have emerged as a promising platform for quantum information science \cite{Redjem.2020.Cassabois,Durand.2021.Dréau,Higginbottom.2022.Simmons} due to their compatibility with scalable photonic integration and manufacturing \cite{psiquantum2025manufacturable}.
Among the silicon color centers, T centers have shown highly coherent ground state electron and nuclear spins \cite{Bergeron.2020.Simmons}. 
With an optical transition in the telecommunication O-band, the T center has become a spin-photon interface candidate suitable for quantum networking and communication applications \cite{Higginbottom.2022.Simmons,Afzal.2024.Yoneda}.

\begin{figure*}[t]
    \centering
    \includegraphics[width=2\columnwidth]{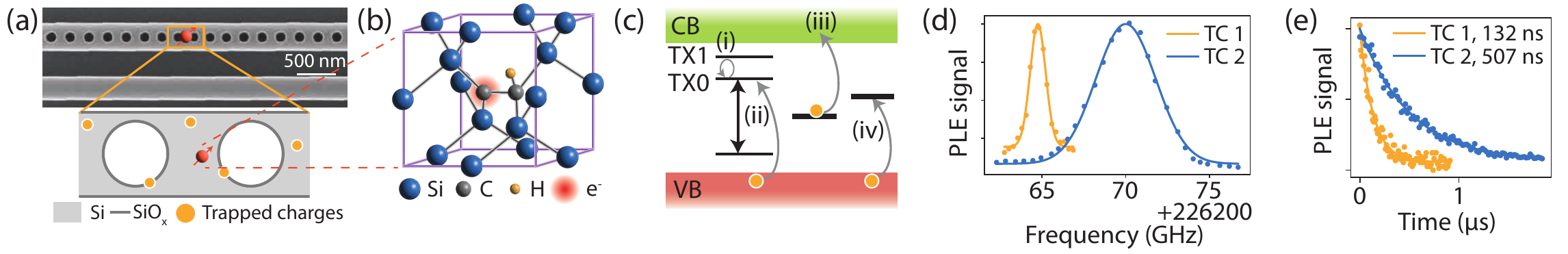}
    \caption{Optical linewidth broadening of T centers. 
    (a) SEM image of the device, a photonic crystal cavity coupled to a waveguide. The zoomed-in schematics illustrate a T center with nearby fluctuating charge traps in the bulk and on the surfaces. 
    (b) Atomic structure of a T center. 
    (c) T center level structure and possible mechanisms leading to linewidth broadening: (i) thermal transition between TX0 and TX1 states, (ii) ionization, (iii-iv) charge transfer between a trap state and (iii) conduction band (CB)  or (iv) valence band (VB).
    (d) The photoluminescence excitation (PLE) spectra and (e) cavity-enhanced lifetimes of two T centers under study.}
    \label{fig1}
\end{figure*}


A major challenge for indistinguishable photon generation with color centers is optical linewidth broadening. 
The linewidth broadening includes homogeneous broadening involving processes faster than the emitter lifetime, and slow spectral diffusion usually due to electric field noise \cite{Beukers.2024.Borregaard,Ngan.2024.Sun}.
For color centers in diamond \cite{Rodgers.2021.Leon,Orphal‐Kobin.2025.Schröder} and silicon carbide \cite{Bader.2024.Castelletto,Lukin.2020.Vučković4,Castelletto.2020.Boretti}, their integration into nanophotonic devices exacerbates the spectral diffusion \cite{Wolters.2013.Benson,Faraon.2012.Beausoleil,Ruf.2019.Hanson,Orphal-Kobin.2023.Schrödera1,Robledo.2010.Hanson} and the laser excitation plays a significant role in introducing the spectral diffusion via reconfiguration of the local charge environment or ionization of the color centers \cite{Stolpe.2025.Taminiau,Orphal-Kobin.2023.Schrödera1,Robledo.2010.Hanson}.
For silicon T centers, photoluminescence excitation (PLE) linewidths as narrow as 33~MHz have been observed in ultra-high-quality bulk silicon at 1.4~K \cite{Bergeron.2020.Simmons}. However, in photonic devices and at higher temperatures, both homogeneous broadening and spectral diffusion contribute to further linewidth broadening.
For example, in typical nanophotonic devices at 4~K, the spectral diffusion (a few GHz) is about an order of magnitude larger than the homogeneous broadening (hundreds of MHz) and four orders of magnitude larger than the lifetime-limited linewidth (a few hundred kHz), creating a large gap between the lifetime-limited linewidth and the measured spectral linewidth \cite{Higginbottom.2022.Simmons}.
Although the homogeneous linewidths of T centers in nanophotonic devices are comparable to the results in bulk silicon \cite{DeAbreu.2023.Simmons}, the causes and timescales of the total linewidth broadening for T centers have not been explored.

Here, we study the homogeneous linewidth and spectral diffusion of T centers in a photonic crystal cavity. 
We use spectral hole burning and tunable T center lifetime via the Purcell effect to investigate the linewidth broadening effects within the T center lifetime. The homogeneous linewidth is unchanged when the lifetime is tuned from 102~ns to 725~ns. 
To study linewidth broadening at longer timescales, we use check-probe spectroscopy \cite{Stolpe.2025.Taminiau} and find that the optical resonance is stable without broadening in the dark up to $3$ ~ms.
We identify the cause of linewidth broadening as laser excitation by inserting laser pulses in the check-probe sequence.
We show that this laser-induced spectral diffusion is not a resonant effect associated with T center excitation.



\noindent{}\textbf{Device and T center PLE measurements.} 
We study the spectral properties of T centers in a one-dimensional photonic crystal cavity shown in Fig.~\ref{fig1}a. 
The thickness of the silicon device layer is 220~nm, and the minimal width of silicon between adjacent holes at the electric field maximum is 172~nm.
Further details of the device, fabrication, and T center creation are discussed in \cite{Komza.2025.Sipahigil}.

The spectral properties of T centers (atomic structure shown in Fig.~\ref{fig1}b) are influenced by the structure of the excited states that form a bound exciton series. The first two excited states, TX0 and TX1, are split by a small energy gap of around 1.76~meV \cite{Bergeron.2020.Simmons} (Fig.~\ref{fig1}c).
This small gap enables rapid thermal transitions between TX0 and TX1 and  introduces a broadening strongly dependent on the local temperature (Fig.~\ref{fig1}c, process i). 
The T center thermal broadening is a rapid process compared to the lifetime of a T center \cite{Bergeron.2020.Simmons} even with Purcell-enhancement \cite{Lee.2023.Waks,Johnston.2023.Chen,Komza.2025.Sipahigil}, thus contributing to the homogeneous linewidth.
Local charge trap reconfiguration and ionization-induced local charge shift are other common sources for spectral diffusion in color centers \cite{Stolpe.2025.Taminiau,Pieplow.2024.Schröder,Ji.2024.Du4xl,Candido.2021.Flatté,Anderson.2019.Awschalom,Delord.2024.Merilesifp,dolde2014nanoscale}. 
For T centers, the ionization (process ii in Fig.~\ref{fig1}c) and charge trap reconfiguration (process iii and iv in Fig.~\ref{fig1}c) happen at longer timescales than thermal broadening, resulting in further broadening of the linewidth. 
We denote the instantaneous frequency of the ground to TX0 optical transition as $\omega_T(t)$. 
We investigate the  frequency fluctuation distribution ($\omega_T(t + \tau)-\omega_T(t )$) at the timescale of $\tau$, and the corresponding linewidth $\gamma(\tau)$.  
Specifically, the spectral diffusion can be modeled as spectral random walk and is known to exhibit linearly increasing linewidth at short times $\gamma(\tau) = \gamma_d \tau$ \cite{zumofen1994spectral,Stolpe.2025.Taminiau}.
We study the diffusion rate $\gamma_d$ under different conditions in our system.

We study T centers in a single cavity where T center 1 (TC1) is measured in the first cooldown and T center 2 (TC2) in the second cooldown after a room temperature thermal cycling. 
During the first/second cooldown, only TC1/TC2 appears within the cavity resonance at 226.27 THz.
We measure the PLE spectrum at an optical power below the power broadening limit. The FWHM linewidths obtained from Gaussian fits are 1.10~GHz and 4.22~GHz, respectively, for TC1 and TC2 (Fig.~\ref{fig1}d). 
With a single-trace averaging time of 1.7~hours, this linewidth represents the long-time linewidth $\gamma (\tau \rightarrow \infty)$. 
We did not observe spectral wandering between traces for single-trace averaging time ranges from minutes to hours. 
The Purcell-enhanced optical lifetime $T_1$ of TC1 and TC2 are extracted to be 132~ns and 507~ns, respectively (Fig.~\ref{fig1}e).

\begin{figure}[h]
    \centering
    \includegraphics[width=\columnwidth]{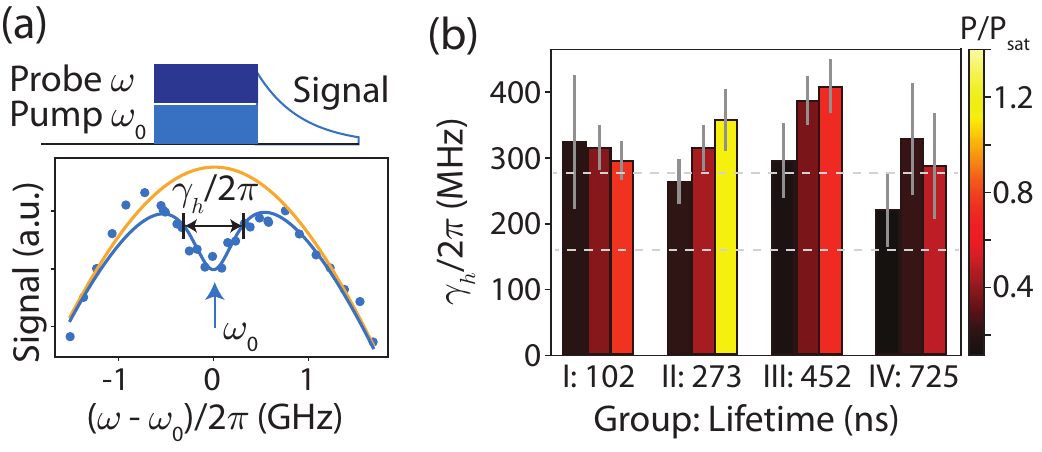}
    \caption{Spectral hole burning. 
    (a) Pulse sequence (upper panel) and a typical hole burning spectrum (lower panel) by sweeping the probe laser frequency.
    (b) Spectral hole width of four lifetime groups with the bar color indicating the total measurement power. The upper dashed line shows the average hole width of the lowest power measurement across the four groups, while the lower dashed line shows the thermal-limited hole width calculated from the cryostat base temperature.}
    \label{fig2}
\end{figure}


\noindent{}\textbf{Probing homogeneous linewidth using spectral hole burning.}
The linewidth broadening within the optical lifetime, i.e., the homogeneous linewidth, can be measured via spectral hole burning \cite{siegman1986lasers,DeAbreu.2023.Simmons}. 
We send a pump laser pulse to TC1 at its center frequency $\omega_0 = \omega_T$ and sweep the frequency of a probe laser pulse $\omega$ across the T center spectral range. 
The two pulses are sent in simultaneously, and the length is set to 1~$\mu$s to reach the steady state before we collect the emission from the T center as the signal (Fig.~\ref{fig2}a top). 
We note that because of the saturation spectroscopy nature of the hole burning measurement, the time scale of the linewidth broadening is set by the lifetime of the emitter instead of the pulse length \cite{siegman1986lasers}.
Due to the saturation of the single-photon emitter, the pump laser excites the emitter when it is close to the pump frequency and suppresses the subsequent excitation from the probe laser.
The saturation effect is absent when the emitter drifts away from the pump laser frequency.
Therefore, this sequence creates a spectral hole for the single emitter (Fig.~\ref{fig2}a bottom). 
The width of the spectral hole $\gamma_h$ is twice the homogeneous optical linewidth of the emitter $\gamma(\tau = T_1)$ at low pump and probe power \cite{siegman1986lasers}.

\begin{figure*}[t]
    \centering
    \includegraphics[width=2\columnwidth]{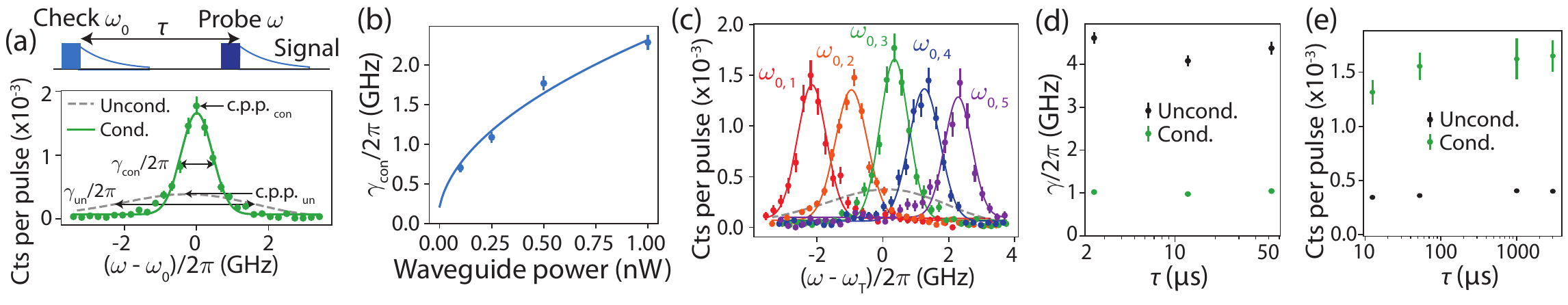}
    \caption{Check-probe spectroscopy.
    (a) Pulse sequence (upper panel) and typical unconditional and conditional spectra (lower panel) from the check-probe spectroscopy. 
    (b) Conditional linewidth as a function of the laser pulse power in the waveguide where check and probe pulses are set at the same power.
    (c) Conditional lineshape at check detunings ($(\omega_{0,j} - \omega_T) / 2\pi$) of (-2.15, -0.96, 0.35, 1.26, 2.28)~GHz for $j$ from 1 to 5. The dashed curve shows the unconditional spectrum.
    (d) Unconditional and conditional linewidths for $\tau$ = 2.5, 12.5, and 52.5 $\mu$s.
    (e) Unconditional and conditional peak counts per pulse for $\tau$ = 12.5, 52.5, 1000, 3000 $\mu$s. }
    \label{fig3}
\end{figure*}

To investigate the linewidth change for $\tau < 1~\mu$s, we control the optical lifetime of TC1 by changing the cavity-TC1 detuning via gas tuning the cavity frequency \cite{Komza.2025.Sipahigil}. 
At each lifetime (102, 273, 452, 725)~ns, we extract $\gamma_h$ from the Lorentzian fit of the hole on a Gaussian background.
To ensure the hole width is in the low power regime, we measure three different total waveguide powers during the pump and probe pulse for each lifetime (Fig.~\ref{fig2}b).
There is no clear trend across different lifetime groups, suggesting that no additional homogeneous broadening from laser-related processes occurs in $102$~ns $< \tau < 725$~ns, even under laser power close to saturation power $P_\textrm{sat}$.

The average $\gamma_h/2\pi$ for the four groups at the lowest waveguide power is 277~MHz, which is larger than the hole width 161~MHz calculated from base temperature $T = 3.41$~K and TX0-TX1 splitting of 1.76~meV \cite{Bergeron.2020.Simmons}. 
The difference indicates the higher local temperature at the sample (App.~\ref{app:hole burning}) or smaller TX0-TX1 splitting.


\noindent{}\textbf{Probing spectral diffusion using check-probe spectroscopy.}
We investigate the timescale of the spectral diffusion, where the optical linewidth further broadens from its homogeneous linewidth by processes slower than the optical lifetime. 
This study employs the check-probe spectroscopy \cite{Stolpe.2025.Taminiau} on TC2.
We send a check laser pulse at frequency $\omega_0$, wait for time $\tau$, and send another probe laser pulse at frequency $\omega$.
We collect the counts from the T center in both the emission window after the check pulse and the one after the probe pulse. 
A click in the check emission window indicates that the T center is close to resonance with $\omega_0$ up to the power-broadened instantaneous linewidth. 
After wait time $\tau$, the T center frequency distribution with possible spectral diffusion in this timescale can be probed by sweeping the probe laser frequency $\omega$. 
An example at $\tau = 2.5\ \mu$s is shown in Fig.~\ref{fig3}a. 
The \textit{unconditional} lineshape is measured by all signal counts per pulse in the probe emission window at different $\omega$.
The \textit{conditional} lineshape is obtained by counting the clicks in the emission window only when there is a click in the check emission window before this probe.
The conditional linewidth $\gamma_\mathrm{con}/2\pi = 1.02$~GHz is narrower than the unconditional PLE linewidth $\gamma_\mathrm{un}/2\pi = 4.62$~GHz. 
Additionally, counts per pulse click rate in the probe window is higher in the case of the conditional lineshape because of the narrower spectral linewidth. 
The observations suggest once the T center frequency is confirmed in the check window, the spectral diffusion in the waiting time is less severe compared to the PLE results.

Before studying check-probe spectroscopy under various conditions, we first explore the broadening effects associated with laser power in the measured conditional linewidth (Fig.~\ref{fig3}b) with $\tau = 2.5\ \mu$s.
These broadening effects include both power broadening and possible laser-induced linewidth broadening.
The check and probe laser pulses are tuned to be at the same power.
Using the power-broadening relation $\gamma_\mathrm{con} = \gamma_\mathrm{con, 0}\sqrt{1 + P/P_\mathrm{sat}}$ where $P_\mathrm{sat}$ is the saturation power and the zero-power conditional linewidth $\gamma_\mathrm{con, 0}$, we obtain $\gamma_\mathrm{con, 0}$ to be $0.2\pm0.4$~GHz. 
Although obtaining an accurate zero-power linewidth will require more measurements at the low-power regime with low count rate, the conditional analysis already reveals the power dependence previously not observable due to the large spectral linewidth of PLE. 
The power-dependence measurement also indicates that the limit of $\gamma_\mathrm{con}$ is mainly set by laser-power-related broadening instead of spectral diffusion within the wait window of $\tau = 2.5\ \mu$s.
In the following measurements, we use the waveguide power of 0.25~nW to reduce the power broadening effect while maintaining a reasonable count rate. 

To confirm that the T center spectral diffusion is around the pinned frequency by the check pulse, we repeated the measurements at different $\omega_0$ around the PLE T center frequency distribution.
The absolute count rate in the check window drops as $\omega_0$ is detuned away from $\omega_T$, requiring longer acquisition time. 
The linewidth and click rate per pulse in the conditional lineshape remains unchanged with the conditional center frequency set by $\omega_0$ (Fig.~\ref{fig3}c). 

Further, we investigate whether the spectral diffusion shows up at long wait time $\tau$. 
For $\tau$ up to 52.5~$\mu$s, we sweep the probe frequency $\omega$ and extract the linewidth (Fig.~\ref{fig3}d), showing that the conditional linewidth does not increase with wait time in this regime.
For longer wait time $\tau$ up to 3~ms, we use the counts per pulse click rate (c.p.p.) at $\omega = \omega_0$ (Fig.~\ref{fig3}e) as a measure of how likely the T center frequency remains stable without drifting away, where the drift will cause a decrease in c.p.p..
The c.p.p.$_\textrm{con}$ shows a slight increase within the errorbar for long wait times, suggesting no observable increase in conditional linewidth. 
In summary, up to $\tau=3$~ms, we do not observe spectral diffusion induced by the dark wait time. 

\begin{figure}[h]
    \centering
    \includegraphics[width=\columnwidth]{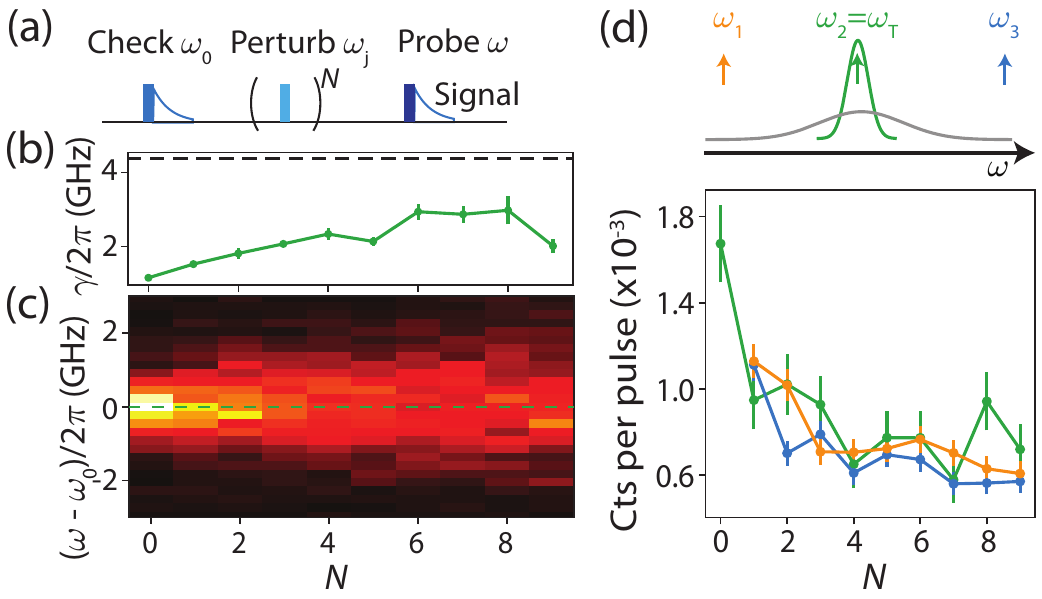}
    \caption{Laser-induced spectral diffusion.
    (a) Pulse sequence with $N$ laser perturb pulses inserted between the check pulse and probe pulse. 
    (b) The linewidth and (c) lineshape of the conditional spectra for $N$ perturb pulses.
    (d) Top: schematics of the three perturb laser frequency relative to the T center frequency and linewidth with the green curve showing the conditional lineshape and the gray curve showing the unconditional lineshape. Bottom: the peak counts per pulse versus $N$ for the three perturb laser frequency with the color corresponding to the arrow color in the top panel.}
    \label{fig4}
\end{figure}

\noindent{}\textbf{Laser-induced spectral diffusion.}
To investigate the cause of spectral diffusion, we modify the check-probe sequence and insert a perturb sequence in the middle. 
The perturb sequence consists of repeated laser pulses (Fig.~\ref{fig4}a), which has been observed to induce spectral diffusion in diamond and SiC \cite{Orphal-Kobin.2023.Schrödera1,Stolpe.2025.Taminiau}. 
First, we set the perturb laser frequency to be resonant with the center frequency of TC2 $\omega_T$, with individual pulse duration of 200~ns and period of 2.5~$\mu$s. 
The check pulse is resonant ($\omega_0=\omega_T$) and we sweep the frequency of the probe pulse to extract the conditional linewidth when clicks happen in both the check and probe windows. 
For perturb pulse number $N$ from 0 (no perturb pulse) to 9, the linewidth increases from 1.16~GHz to 2.98~GHz (Fig.~\ref{fig4}b), approaching the PLE linewidth. 
The corresponding line shapes (Fig.~\ref{fig4}c) change gradually from peaked to flattened with increasing $N$.
At laser pulse number $N \leq 4$, the conditional linewidth $\gamma/2\pi$ increases linearly with the laser pulse at a rate of 1.45 GHz per microsecond of laser pulse (App.~\ref{app:rate}). 
The fluctuation of the linewidth at large $N$ results from the photon shot noise as the line shape flattens and the center frequencies accumulate less counts than small $N$ results. 
Additionally, we observe an increase of linewidth with increased perturb pulse power when keeping the duration and number of the perturb pulses (App.~\ref{app:Check-probe}).

This laser-induced spectral diffusion could be due to the local charge environment change from either resonant effects such as the ionization of TC2 itself (ii in Fig.~\ref{fig1}c) or the reconfiguration of nearby charge traps (iii and iv in Fig.~\ref{fig1}c).
To distinguish these two effects, we repeat the measurements with off-resonant perturb pulses. 
We choose the perturb pulse frequency to be either -5.1~GHz or 5.95~GHz detuned from TC2 center frequency $\omega_T$.
These detunings are large enough to avoid the excitation of TC2 (illustrated in Fig.~\ref{fig4}d), especially after its frequency is heralded by the check pulse. 
On the other hand, these detunings are small compared to the cavity linewidth of 17.9~GHz, enabling effective coupling of photons into the cavity. 
Similar to the previous section, we use the peak counts per pulse (c.p.p.) as a measure of the height of the line shape, which is inversely proportional to the linewidth. 
The resonant c.p.p. extracted directly from Fig.~\ref{fig4}c is used as a reference. 
The perturbation with different detunings follow the same trend as the resonant perturbation (Fig.~\ref{fig4}d), indicating the resonant effects including the T center ionization is not the major source for laser-induced spectral diffusion for T centers.


\noindent{}\textbf{Conclusion and discussion.}
Our spectral hole burning and check-probe spectroscopy results reveal possible origins and the relevant timescales of T center linewidth broadening in nanophotonic devices.
The observations suggest that at short timescales below hundreds of nanoseconds, homogeneous linewidth does not increase with optical lifetime in current experiments.
Without a laser perturbation, the T center in the dark can preserve its frequency for at least 3~ms.
Laser perturbation, even with frequencies below the silicon bandgap, plays a major role in introducing spectral broadening.
The broadband nature of the effect suggests that broadening is due to the reconfiguration of nearby charge environment instead of the ionization of the T center itself.

These conclusions can be used to design more efficient pulse sequences for indistinguishable photon generation with T centers. 
As an example, the entanglement generation between two T centers using emitted photon interference \cite{Pompili.2021.Hanson,Hermans.2022.Hanson,Robledo.2010.Hanson,Bernien.2013.Hanson,Brevoord.2024.Hanson,Abobeih.2022.Taminiau} requires frequency matching of these two emitters.
Instead of waiting for the two emitters to arrive at the same frequency among the frequency jitters, each emitter can be prepared independently using the check pulse with a click heralding the successful preparation at the laser frequency \cite{Brevoord.2024.Hanson,Bernien.2013.Hanson,robledo2011high}.
The emitter prepared first can preserve its frequency in the dark while waiting for the other emitter to be ready. 
This heralded preparation method avoids repeating the full sequence numerous times. 
Instead, the repeat-until-success part only happens for the short emitter excitation sequence.
On the other hand, the spectral diffusion dominated by laser-induced charge reconfiguration may pose a challenge to preserving the emitter frequency during spin initialization using optical pumping. 
Detailed studies considering specific power and duration of the optical pumping pulse will be necessary to determine whether the optical transition frequency can be prepared before the spin initialization.

This study also brings up questions for further exploration. 
Carrying out the check-probe spectroscopy in a waveguide \cite{DeAbreu.2023.Simmons,Islam.2024.Waks} instead of the photonic crystal cavity could provide broader-band information about the charge trap response in silicon.
Further study is needed to investigate the origin and dominant location of the charge traps, which can be broadly categorized as bulk charge traps and surface charge traps. 
Some surface charge traps at the Si/SiO$_2$ interfaces such as the $P_b$ centers are well-studied \cite{stirling2000dangling,boehme2005pulsed,kato2006origin}.
Using surface passivation techniques \cite{brower1988kinetics} or changing the nanophotonic structure dimension will help to separate the contribution of different sources, which could lead to solutions to reduce the spectral diffusion.
The current check-probe spectroscopy linewidth is limited by laser-power-related broadening. 
Using Hong-Ou-Mandel interference \cite{Hong.1987.Mandel,Santori.2002.Yamamoto,Gazzano.2013.Senellart,Komza.2024.Sipahigil} enables the access to power-broadening-free linewidth at various wait time. 
Lastly, cooling the device to 1~K will exponentially lower the thermal-limited linewidth \cite{Bergeron.2020.Simmons,DeAbreu.2023.Simmons}.
The lower temperature could unveil frequency jitter mechanisms currently masked by the thermal linewidth. 

\noindent{}\textbf{Data availability.}
The data that support the findings of this study are openly available in Zenodo \cite{zhang_dataset_2025}.

\noindent{}\textbf{Note added.}
During the preparation of this manuscript, we became aware of the related work by Bowness and Meynell \textit{et al}. \cite{bowness2025laser}

\noindent{}\textbf{Acknowledgments.}
We thank Zi-Huai Zhang and Mi Lei for helpful discussions and feedback on the manuscript. This work was primarily supported by the Office of Advanced Scientific Computing Research (ASCR), Office of Science, U.S. Department of Energy, under Contract No. DE-AC02-05CH11231 and Berkeley Lab FWP FP00013429. L.K. and A.S acknowledge support from the NSF (QLCI program through grant number OMA-2016245, and  Award No. 2137645). X.Z. acknowledges support from the Miller Institute for Basic Research in Science. The devices used in this work were fabricated at the Berkeley Marvell NanoLab.


\bibliography{references}

\clearpage

\appendix
\renewcommand{\thefigure}{S\arabic{figure}}
\setcounter{figure}{0}

\section{Spectral hole burning: setup and temperature dependence}
\label{app:hole burning}

\begin{figure*}[t!]
 	\centering
 	\includegraphics[width=2\columnwidth]{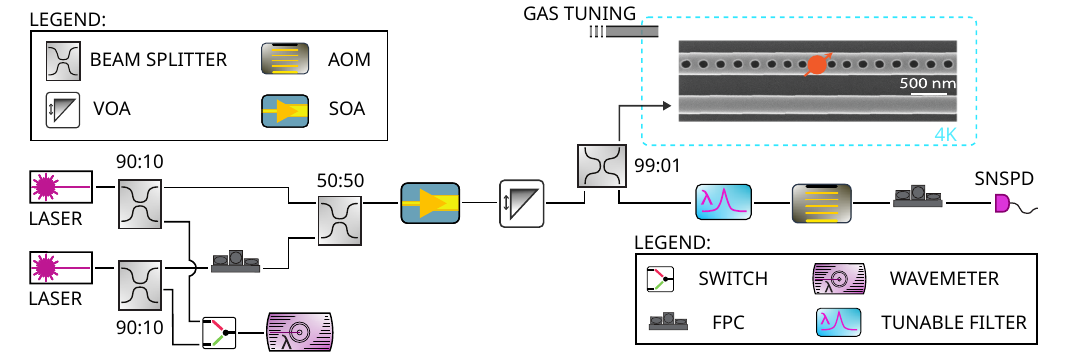}
 	\caption{Setup for the spectral hole burning. AOM: acousto-optic modulator, SOA: semiconductor optical amplifier, FPC: fiber polarization controller.}
\label{figS1}
\end{figure*}

We use the setup depicted in Fig.~\ref{figS1} for the spectral hole burning. Two lasers (Santec TLS-570 and Instatune FP4209) are locked using a wavemeter (High Finesse WS/6-200) with a PID loop. After getting combined at the 50:50 BS, a semiconductor optical amplifier (SOA, Aerodiode SOA-3PI, 1 ns rise time) creates the optical pulses and a variable optical attenuator (VOA, Agiltron)  controls the pulse power. After the FPC, which is used to align the polarization to the optical waveguide, the laser pulses are routed to the device inside a 4K cryostat (Montana S200) using a 99:01 BS. The coupling to the device is done with a lensed fiber (OZ Optics) mounted on nano-positioners (Attocube ANPx101/LT and ANPz102/LT). The light reflected from the device and emitted from the T-center is spectrally filtered using a tunable bandpass filter (WL photonics, flat top with bandwidth of~\SI{0.5}{nm}). The strong control light is time-gated using an AOM (Aerodiode AOM model 10, rising time 10 ns). A final FPC aligns the polarization to maximize the T-center signal on the SNSPD (Quantum Opus, quantum efficiency ~60\%). TTL control pulses are generated with Swabian Instruments Pulse Streamer 8/2, and SNSPD clicks are recorded with a Swabian Instruments Time Tagger Ultra. 

A steel hose (outer diameter 1/16") is used to gas-tune the cavity to the T-center resonance. Since it can only redshift the cavity, we use the laser on resonance with the cavity to blueshift it and maintain the resonance condition (see~\cite{Komza.2025.Sipahigil} for more details).

\begin{figure*}[t!]
 	\centering
 	\includegraphics[width=0.75\columnwidth]{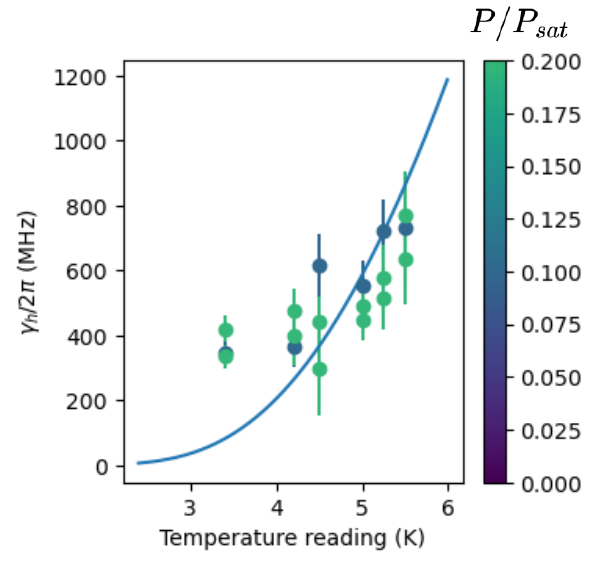}
 	\caption{Spectral hole burning linewidth with temperature of the cryostat. Spectral hole burning linewidth for several temperatures of the cryostat: from base temperature of 3.4 K to 5.5K. The colors of the datapoints represent different powers relative to the saturation power $P_{sat}$. The blue curve is the expected thermal linewidth assuming TX0-TX1 splitting of 1.76~meV. All errorbars are one standard deviation.}
\label{figS2}
\end{figure*}

We measured the spectral hole burning for several chamber temperatures, using the same sequence as in Fig.~\ref{fig2} in the main text. We report the results in Fig.~\ref{figS2} for several optical powers (relative to the saturation power). We note that the trend of the linewidth follows quite closely the expected trend (blue solid line, from the thermal broadening~\cite{Bergeron.2020.Simmons}) for higher chamber temperature (up to 5.5 K). For lower temperatures, we note a plateau around the hole width of 400~MHz, indicating a higher effective temperature of the device than the base temperature of the cryostat.
This set of data is taken in a separate cool down from the main text Fig.~\ref{fig2}b, which may result in an increased device temperature and thus a larger hole width.

\section{Check-probe spectroscopy: setup and power dependence}
\label{app:Check-probe}

\begin{figure*}[t!]
 	\centering
 	\includegraphics[width=2\columnwidth]{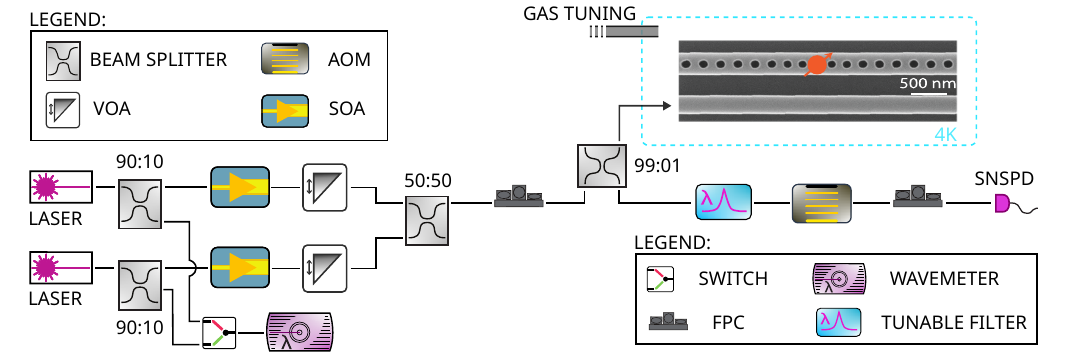}
 	\caption{Setup for the check-probe spectroscopy. VOA: variable optical attenuator, AOM: acousto-optic modulator, SOA: semiconductor optical amplifier, FPC: fiber polarization controller. }
\label{figS3}
\end{figure*}

We use the setup depicted in Fig.~\ref{figS3} for the check-probe spectroscopy. The setup is similar to the one of the spectral hole burning. We describe only the differences here. Each laser is followed by their dedicated SOA (Aerodiode SOA-3PI, 1 ns rise time) and VOA (Agiltron MSOA) to independently create optical pulses and set the power (measured at the unused arm of the 50:50 BS with a power meter - Thorlabs S154C). Note that the high on-off ratio of the SOAs allows for a small background in the measurements reported in Fig.~\ref{fig3}, enabling a negligible amount of spurious coincidences. After the combination at the 50:50 BS and the FPC the setup is identical to the one described in~\ref{app:hole burning}.

\begin{figure*}[t!]
 	\centering
 	\includegraphics[width=0.75\columnwidth]{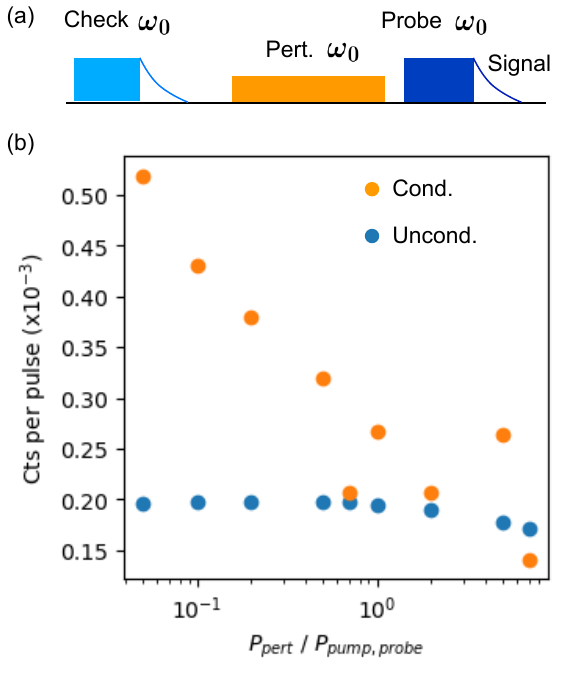}
 	\caption{Power dependence of laser-induced spectral diffusion. (a) Pulse sequence: a check-probe scheme is used with a long perturbation pulse in between (labeled 'Pert.'). (b) Conditional (orange) and unconditional (blue) count per pulse as a function of perturbation laser power. }
\label{figS4}
\end{figure*}

Similar to Fig.~\ref{fig4}, we use a check-probe scheme to assess the laser-induced spectral diffusion caused by a long laser pulse. In Fig.~\ref{figS4}a, the pulse sequence consists of a check and a probe pulse (locked at the center frequency $\omega_0$, pulse length of 200 ns, shown in blue), and an 8-$\mu s$ long perturbation pulse in between the check and probe pulse (locked at $\omega_0$, shown in orange). The single photon arrival time is represented by the blue line with exponential decay labeled 'signal'. We use a fixed pump-probe power of 2.5\% of the saturation power.

Since a full scan of the laser wavelength across the T center PLE span would be too time-consuming, we use, as done for Fig.~\ref{fig3}e, the count per pulse (c.p.p.) at the center frequency to infer the linewidth. We report the results in Fig.~\ref{figS4}b. As in the main text, we calculate the conditional and unconditional c.p.p. to compare the influence of the perturbation pulse . Since the conditional c.p.p. decreases with increasing power of the perturbation pulse, we can infer an increase in the spectral diffusion. 

The perturbation pulse mimics the initialization step of the T center spin using optical pumping. We compare the laser power of the perturbation pulse to the power we expect for optical pumping of the spin \cite{Higginbottom.2022.Simmons}.
Although the actual optical pumping power needs to be calibrated for the specific setup, it is likely the optical pumping will lead to a considerable increase in spectral diffusion. Therefore, it could be challenging to herald with an initial pulse and maintain the resonance frequency through the spin initialization process via optical pumping. More careful calibration and pulse sequence design will be necessary in this case. 

\section{Spectral diffusion rate}
\label{app:rate}

The spectral diffusion from charge trap reconfiguration has been modeled using random walk theory \cite{zumofen1994spectral,Stolpe.2025.Taminiau}. 
The model suggests that the inhomogeneous linewidth is determined by a Lorentzian propagator capturing the spectral diffusion and its homogeneous linewidth.
The linewidth of the Lorentzian propagator grows linearly with time at the beginning, resulting in a linear growth of the spectral-diffusion dominated inhomogeneous linewidth $\gamma = \gamma_d \tau$ where $\gamma_d$ is the spectral diffusion rate. 

\begin{figure*}[t!]
 	\centering
 	\includegraphics[width=1\columnwidth]{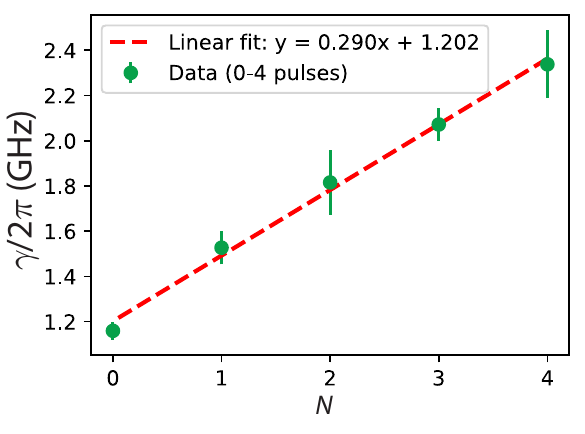}
 	\caption{Linear fitting of the conditional linewidth in Fig.~\ref{fig4}b at laser pulse number 0-4.}
\label{figS5}
\end{figure*}

By fitting the conditional linewidth under the first 4 perturb pulses in Fig.~\ref{fig4}b, we obtain the spectral diffusion of 0.29 GHz per laser pulse (Fig.~\ref{figS5}).
Given 200 ns of laser pulse length, the spectral diffusion rate under 0.25 nW of waveguide power is $\gamma_d/2\pi = 1.45$ GHz$/\mu$s.

\end{document}